\documentclass[12pt]{article}
\usepackage[square]{natbib}
\usepackage{a4}
\usepackage{latexsym} 
\usepackage{amssymb}  
\usepackage{graphicx}
\usepackage{epsfig}
\usepackage{amsmath}

\newcounter{zyxabstract}     
\newcounter{zyxrefers}        

\newcommand{\newabstract}
{\newpage\stepcounter{zyxabstract}\setcounter{equation}{0}
\setcounter{footnote}{0}}

\newcommand{\rlabel}[1]{\label{zyx\arabic{zyxabstract}#1}}
\newcommand{\rref}[1]{\ref{zyx\arabic{zyxabstract}#1}}

{\section*{References}\setcounter{zyxrefers}{0}
\begin{list}
{[\arabic{zyxrefers}]}
{\usecounter{zyxrefers}\setlength{\parindent}{0cm}\setlength{\itemsep}{0cm}}

}
{\end{list}}
\newenvironment{thebibliographynotitle}[1] 
{\section*{References}\setcounter{zyxrefers}{0}
\begin{list}{[\arabic{zyxrefers}]}
{\usecounter{zyxrefers}\setlength{\parindent}{0cm}\setlength{\itemsep}{-1.5mm}}}
{\end{list}}

\renewcommand{\bibitem}[1]{\item\rlabel{y#1}}
\renewcommand{\cite}[1]{[\rref{y#1}]}      

\newcommand{\OP}[1]{\check{#1}}

\newcommand{\VC}[1]{\boldsymbol{#1}}

\newcommand{\BA}[1]{\langle #1 \mid}

\newcommand{\KT}[1]{\mid #1 \rangle}

\newcommand{\bibentry}[5]{#1, {#2} {\bf #3}, (#5) #4}

\begin{document}
\begin{titlepage}
\begin{flushright}
{\small
}
\end{flushright}

\begin{center}
{\LARGE\bf Workshop on Chiral Forces \newline in Low Energy Nuclear Physics \newline - the LENPIC Meeting$^{*}$~~~~~}
\\[1cm]
Jagiellonian University, Krak\'ow, Poland\\
February 10-11, 2017\\[1cm]
{\bf Jacek Golak,} {\bf Roman Skibi\'nski}
\\[0.3cm]
M. Smoluchowski Institute of Physics, Jagiellonian University, \\
30-384 Krak\'ow, Poland\\[1cm]
{\large ABSTRACT}
\end{center}
These are the proceedings of the international workshop on "Chiral Forces
in Low Energy Nuclear Physics - the LENPIC Meeting"
held at the Jagiellonian University, Krak\'ow, Poland
from February 10 to 11, 2017. The workshop focused on the new generation of chiral
forces with the semi-local regularization and their applications to few- and many-nucleon systems.
Each talk is represented by a short contribution.

\vfill
\noindent\rule{6cm}{0.3pt}\\
\footnotesize{$^*$ This workshop 
is a part of the LENPIC project activity and 
was supported by the Polish National Science Center under Grants No. DEC-2013/10/M/ST2/00420
and by the Faculty of Physics, Astronomy and Applied Computer Science of Jagiellonian University
within the KNOW project.
}
\end{titlepage}

\tableofcontents

\newpage

\addcontentsline{toc}{section}{Introduction}
\section*{Introduction}

\hskip 0.6 true cm
Chiral effective field theory (EFT) provides a powerful
framework for ana\-ly\-zing low-energy nuclear structure and
reactions in full agreement with the symmetries of quantum chromodynamics. 
It allows one to derive nuclear forces and currents
in a systematically improvable way, using the so-called
chiral expansion.

The chiral approach provides a natural explanation of
the observed hierarchy of nuclear forces.
In particular chiral EFT is expected to provide an ultimate
theoretical solution of the long-standing three-nucleon force
problem. This means that the chiral interactions are very
attractive input for ab-initio studies of both nuclear structure
and reactions. The recently developed new generation
of the chiral potentials with semi-local regularization
gives us hope for overcoming  technical obstacles present
for the older models of chiral interactions.

LENPIC is a worldwide collaboration that aims to develop
chiral effective field theory nucleon-nucleon and many-nucleon
interactions complete through fifth order in the chiral
expansion. Using these new interactions, LENPIC intends to
solve the structure and reactions of light nuclei including
electroweak observables with consistent treatment of
the corresponding single and many-nucleon currents.

LENPIC brings together scientists from the following institutions:
Ruhr-University Bochum, Germany,
University of Bonn, Germany,
Technical University of Darmstadt, Germany,
Jagiellonian University, Krakow, Poland,
Iowa-State University, USA,
J\"ulich Research Centre, Germany,
Kyushu Institute of Technology, Japan,
Ohio State University, USA,
Orsay Institute of Nuclear Physics, France,
and TRIUMF, Canada.
More information can be found at the LENPIC website: www.lenpic.org .

This workshop took place at the Faculty of Physics, Astronomy and Applied Computer Science
of the Jagiellonian University, Krak\'ow, Poland, in February 2017.
We would like to thank all the sponsors and the participants for making that meeting
an exciting and lively event.

\vspace{0.5cm}

\hfill Jacek Golak and Roman Skibi\'nski

\newpage

\addcontentsline{toc}{section}{E.~Epelbaum: ${\rm LENPIC:~Status,~Challenges~and~Perspectives}$}
\newabstract 
\begin{center}
{\large\bf LENPIC: Status, Challenges and Perspectives}\\[0.5cm]
{\bf Evgeny Epelbaum}$^1$  \\[0.3cm]
$^1$Institut f\"ur Theoretische Physik II, Ruhr-Universit\"at Bochum,\\
44780 Bochum, Germany\\[0.3cm]
\end{center}

Chiral effective field theory (EFT) has become a standard tool to analyze
low-energy reactions involving pions, nucleons and external
electroweak sources. Most of the ab-initio calculations of light and
medium-mass nuclei and nuclear reactions are nowadays performed
based on the Hamiltonian and currents derived in chiral EFT. The Low Energy
Nuclear Physics International Collaboration (LENPIC) aims to develop
chiral nuclear forces and currents complete through (at least) fourth
order (i.e.~N$^3$LO) in the chiral expansion and to perform precision ab-initio
studies of the structure and reactions of light and medium-mass
nuclei \cite{LENPIC}.

Recently, we have presented a new generation of nucleon-nucleon (NN)
potentials up to fifth order (i.e.~N$^4$LO) in the chiral expansion
\cite{Epelbaum:2014efa}. These interactions employ
a local coordinate-space regularization of the long-range interaction,
which preserves the analytic structure of the amplitude and
efficiently suppresses unphysical short-range components in the
two-pion exchange making
the spectral-function regularization obsolete. With all relevant $\pi
N$ low-energy constants (LECs) taken from $\pi N$ scattering
without any fine tuning, we were able to demonstrate a clear evidence
of the chiral two-pion exchange. These new potentials, coupled with
the novel approach to estimate truncation error formulated in
\cite{Epelbaum:2014efa}, provide a solid basis for applications within
LENPIC. We found
promising results by applying these novel NN forces beyond the
two-nucleon system to calculate nucleon-deuteron scattering and
selected properties of light nuclei \cite{Binder:2015mbz}, electroweak
NN and 3N reactions \cite{Skibinski:2016dve} and equations of state of
nuclear matter \cite{Hu:2016nkw}. In all cases considered so far, the
observed discrepancies between the theoretical predictions and
experimental data are in a good agreement with the size of
the three-nucleon force (3NF) effects expected in Weinberg's power
counting scheme. Still, more work is needed to further test and/or
adjust the algorithm for estimating uncertainty from the truncation of
the chiral expansion. This applies, in particular, to the determination of the
relevant expansion parameter in calculations of bound- and
excited-state properties.

Clearly, the next step is the inclusion of the 3NF. The expressions
for the three- and four-nucleon forces are available through N$^3$LO,
see \cite{Epelbaum:2015pfa} for a review, and work is
in progress towards completing the derivation  of the remaining
N$^4$LO contributions.
The numerical implementation of the 3NF is challenging and represents
one of the central tasks of the LENPIC Collaboration. Specifically,
the 3NF needs  to be regularized in the way consistent with the NN
potential and expressed in the partial-wave basis. Partial wave
decomposition of a general 3NF can be performed numerically
\cite{Golak:2009ri}, and its computational cost can be further
reduced in the case of local 3NFs \cite{Hebeler:2015wxa}. Work is in
progress towards the numerical implementation of the
regularization of the long-range parts of the 3NF at N$^3$LO which
represents the major challenge. To ensure that the results for 3NF matrix
elements are numerically stable we follow  different computational
strategies by
performing calculations independently in coordinate and momentum
spaces. First benchmarks at the level of the
N$^2$LO 3NF have already been successively performed, and the implementation
of the N$^3$LO terms is in progress.

In parallel to these developments, we have also worked out the
corresponding electromagnetic, axial and
pseudoscalar nuclear currents up to N$^3$LO
\cite{Kolling:2009iq},~\cite{Krebs:2016rqz}, see also
\cite{Piarulli:2012bn},~\cite{Baroni:2016xll}
and references therein for a related work by the JLab-Pisa group
(whose results differ from ours).  Similarly to the 3NFs, the
resulting exchange currents need to be regularized and partial-wave
decomposed. Special care is required to ensure that the symmetries,
which manifest themselves in the form of the continuity equations, are
not destroyed by regularization. This issue is particularly important for
electroweak processes such as e.g.~$^3$H $\beta$-decay and
$\mu$-capture on $^2$H, $^3$H and $^3$He, for which parameter-free
predictions can be made at the level of N$^3$LO (once the strength of the 3NF
LEC $c_D$ is fixed in the strong sector).

In summary, the outline developments within the LENPIC Collaboration
towards the implementation of the 3NFs and current operators derived
in the framework of chiral EFT open the way for precision calculations of light and
medium-mass nuclei with quantified theoretical uncertainties.

\bigskip
It is a pleasure to thank the whole LENPIC for enjoyable collaboration
and the Krak\'{o}w group for the excellent organization of the meeting.

\setlength{\bibsep}{0.0em}
\begin{thebibliographynotitle}{99}

\bibitem{LENPIC}
LENPIC Collaboration, www.lenpic.org

\bibitem{Epelbaum:2014efa}
  E.~Epelbaum, H.~Krebs, U.-G.~Mei{\ss}ner,
  Eur.\ Phys.\ J.\ A {\bf 51}, no. 5, 53 (2015); Phys.\ Rev.\ Lett.\  {\bf 115}, no. 12, 122301 (2015).

\bibitem{Binder:2015mbz}
  S.~Binder {\it et al.} [LENPIC Collab.],
  Phys.\ Rev.\ C {\bf 93}, no. 4, 044002 (2016).

\bibitem{Skibinski:2016dve}
  R.~Skibi\'{n}ski {\it et al.},
  Phys.\ Rev.\ C {\bf 93}, no. 6, 064002 (2016).

\bibitem{Hu:2016nkw}
  J.~Hu {\it et al.},
  arXiv:1612.05433 [nucl-th].

\bibitem{Epelbaum:2015pfa}
  E.~Epelbaum,
  PoS CD {\bf 15}, 014 (2016).

\bibitem{Golak:2009ri}
  J.~Golak {\it et al.},
  Eur.\ Phys.\ J.\ A {\bf 43}, 241 (2010).

\bibitem{Hebeler:2015wxa}
  K.~Hebeler {\it et al.},
  Phys.\ Rev.\ C {\bf 91}, no. 4, 044001 (2015).

\bibitem{Kolling:2009iq}
  S.~K\"olling, E.~Epelbaum, H.~Krebs and U.-G.~Mei{\ss}ner,
  Phys.\ Rev.\ C {\bf 80}, 045502 (2009); Phys.\ Rev.\ C {\bf 84}, 054008 (2011).

\bibitem{Krebs:2016rqz}
  H.~Krebs, E.~Epelbaum and U.-G.~Mei{\ss}ner,
  Annals Phys.\  {\bf 378}, 317 (2017).

\bibitem{Piarulli:2012bn}
  M.~Piarulli {\it et al.},
  Phys.\ Rev.\ C {\bf 87}, no. 1, 014006 (2013).

\bibitem{Baroni:2016xll}
  A.~Baroni {\it et al.},
  Phys.\ Rev.\ C {\bf 94}, no. 2, 024003 (2016).
\end{thebibliographynotitle}

\addcontentsline{toc}{section}{K.~Hebeler: ${\rm Calculation~of~semilocal~3N~interactions~up~to~N3LO:}$ \protect\newline 
                              ${\rm ~~~~~~~~~~~~~status~update~and~recent~developments}$}
\newabstract 
\begin{center}
{\large\bf Calculation of semilocal 3N interactions up to N3LO: status update and recent developments}\\[0.5cm]
{\bf Kai Hebeler}$^{1,2}$\\[0.3cm]
$^1$Institut f\"ur Kernphysik, Technische Universit\"at Darmstadt, 64289 Darmstadt, Germany\\
$^2$ExtreMe Matter Institute EMMI, GSI Helmholtzzentrum f\"ur Schwerionenforschung GmbH, 64291 Darmstadt, Germany\\
\end{center}

During the recent years there has been considerable effort to derive novel
nuclear interactions within chiral EFT. In particular, in
Refs.~\cite{Epelbaum:2014efa},\cite{Epelbaum:2014sza} a novel way of regularizing
nuclear forces was proposed. In contrast to previous nuclear forces, which
employed a non-local momentum-space regulator, the new way of regularizing the long-
range parts of the interaction in a local way leads to significantly reduced
cutoff artifacts and also preserve the analytic structure of the scattering
matrix close to pion threshold. First calculations for light nuclei based on
these new NN forces  are very promising~\cite{Binder:2015mbz}.

For consistent structure and reaction calculations up to N$^2$LO and N$^3$LO
it will be crucial to also include contributions from three-nucleon (3N)
interactions. Recently we  made significant progress in generalizing the new
semi-local regularization scheme to 3N forces. Using the new efficient
framework for calculating unregularized 3N interactions presented in
Ref.~\cite{N3LO} we implement the local coordinate-space regularization
$V_{\rm reg} (\mathbf{r}_{12}, \mathbf{r}_{23}, \mathbf{r}_{13}) = R (r_{12}, r_{23}, r_{13}) V(\mathbf{r}_{12}, \mathbf{r}_{23}, \mathbf{r}_{13})$
in form of convolution integrals in momentum space:
\begin{equation} 
\left< \mathbf{p}' \mathbf{q}' | V_{\rm reg} | \mathbf{p} \mathbf{q} \right> = 
\int 
\frac{d \bar{\mathbf{p}}}{(2 \pi)^3} 
\frac{d \bar{\mathbf{q}}}{(2 \pi)^3} 
\left< \mathbf{p}' \mathbf{q}' | R | \bar{\mathbf{p}} \bar{\mathbf{q}} \right> 
\left< \bar{\mathbf{p}} \bar{\mathbf{q}} | V | \mathbf{p} \mathbf{q} \right> .
\end{equation}

Here $\mathbf{r}_{ij}$ represent the relative distances of the three
particles, $R$ is the regulator function and $\mathbf{p}$ and $\mathbf{q}$ ($\mathbf{p'}$ and $\mathbf{q'}$) are
the Jacobi momenta of the initial (final) state, respectively. Recently we
resolved the numerical challenges related to the practical numerical
calculation of the integrals and have now finished the calculation of all
partial-wave matrix elements at N$^2$LO. Next we will explore their effects in
few- and many-body calculations and then compute all matrix elements up to
N$^3$LO.

\vspace{-0.5cm}
\setlength{\bibsep}{0.0em}
\begin{thebibliographynotitle}{99}
\bibitem{Epelbaum:2014efa}E.\ Epelbaum, H.\ Krebs and U.-G.\ Mei{\ss}ner,
Eur.\ Phys.\ J.\ A {\bf 51}, (2015) 53.
\bibitem{Epelbaum:2014sza}E.\ Epelbaum, H.\ Krebs and U.-G.\ Mei{\ss}ner,
Phys.\ Rev.\ Lett.\  {\bf 115}, (2015) 122301.
\bibitem{Binder:2015mbz} S.\ Binder {\it et al.}, Phys.\ Rev.\ C {\bf 93}, (2016) 044002.
\bibitem{N3LO}K.\ Hebeler, H.\ Krebs, E.\ Epelbaum, J.\ Golak and R.\ Skibinski,
Phys. Rev. C {\bf 91} (2015) 044001.
\end{thebibliographynotitle}

\addcontentsline{toc}{section}{R.~Roth: ${\rm Ab~Initio~Calculations~of~Light~Nuclei~for~Constraining}$ \protect\newline 
${\rm ~~~~~~~~~~Chiral~NN{+}3N~Interactions}$}
\newabstract 
\begin{center}
{\large\bf Ab Initio Calculations of Light Nuclei\\ for Constraining Chiral NN+3N Interactions}\\[0.5cm]
{\bf Robert Roth}$^1$ \\[0.3cm]
$^1$Institut f\"ur Kernphysik, Technische Universit\"at Darmstadt,\\
Schlossgartenstr. 2, 64289 Darmstadt, Germany\\[0.3cm]
\end{center}

The ab initio description of light and medium-mass nuclei starting from two- and three-nucleon interactions derived within 
chiral effective field theory is one of the most dynamic fields in nuclear theory today. 
For light nuclei, i.e., the p-shell and selected sd-shell nuclei, the no-core shell model (NCSM) 
with importance truncation provides access to all relevant nuclei 
and observables \cite{ncsm1},\cite{ncsm2},\cite{itncsm1},\cite{itncsm2}, including ground and 
excited-state energies, radii, density and momentum distributions, and all electromagnetic and weak observables. The sole limitation and source of uncertainty in the NCSM is the convergence of the observables with increasing size of the many-body model space. In order to enhance this convergence, we routinely use similarity renormalization group (SRG) transformations of the Hamiltonian and all relevant observables up to the three-nucleon level \cite{srg}. In combination, SRG-evolved operators and the NCSM provide a very efficient, robust, and universal tool to study ground-state and spectroscopic observables in all p-shell nuclei.

This is an ideal framework for characterizing and constraining next-generation chiral interactions beyond the few-body domain. Traditionally, few-body calculations for bound-state and scattering observables are being used to fit and test new NN+3N interactions. Ab initio NCSM calculations of p-shell nuclei access a new domain of observables that are highly sensitive to different aspects of the input interactions and, thus, provide important additional information on the performance of the interaction that is not accessible in the few-body sector. We can, e.g., probe the spin-orbit and tensor structure through excitation spectra and spectroscopy for states with larger angular momenta, the isospin dependence of the interaction through isospin chains, and the saturation properties through ground-state energies and radii in the upper p-shell.

In order to use p-shell nuclei for testing and constraining the consistent NN+3N interactions 
up to N3LO constructed within the LENPIC collaboration \cite{lenpicNN},\cite{lenpic3N},\cite{lenpicfew}, 
we first need to identify a set of observables that depend sensitively on details of the interaction. In addition to sensitivity to the interaction, these observables should be easily accessible in a consistent way starting from the chiral EFT inputs. This makes ground-state and excitation energies particularly well suited, since they show a good convergence and need no additional input from chiral EFT, unlike electroweak observables that require two-body current contributions consistent with the interaction. Therefore, we explored ground-state and excitation energies of low-lying states with a range of different nuclei using a variety of previous-generation chiral NN+3N interactions. Using $^6$Li and $^{12}$C as examples, we discussed results on the sensitivities of excitation energies \cite{sensitivity}. With preliminary versions of the consistent chiral NN+3N interactions at N2LO with semi-local regulators, we started to explore the dependence of the same excitation energies on the three-body low-energy constant (LEC) $c_D$, fixing the corresponding $c_E$ from a fit to the triton ground-state energy. These exploratory calculations confirm the picture of the previous survey with other interactions \cite{sensitivity} regarding the sensitivity of individual excited states on the interaction.

Working towards the final versions of the LENPIC NN+3N interactions up to N3LO, we propose to use ground-state and excitation energies of p-shell nuclei as additional diagnostic during the fit of the three-body LECs. Though the eventual fit will preferentially focus on few-body observables, monitoring these many-body observables will help to to understand their dependence on individual LECs and the impact of the regulator scheme and scale. Furthermore, many-body observables can provide additional guidance for the choice of LECs, particularly if the few-body observables within their uncertainties do not constrain the LECs sufficiently. We advocate a comprehensive consideration of few- and many-body observables including their theoretical uncertainties when determining optimal parameter ranges for the LECs.

\setlength{\bibsep}{0.0em}
\begin{thebibliographynotitle}{99}
\bibitem{ncsm1}
P. Navrátil, S. Quaglioni, I. Stetcu, B. Barrett, J. Phys. G 36, 083101 (2009).
\bibitem{ncsm2}
B. R. Barrett, P. Navrátil, J. P. Vary, Prog. Part. Nucl. Phys. 69, 131 (2013).
\bibitem{itncsm1}
R. Roth, Phys. Rev. C 79, 064324 (2009).
\bibitem{itncsm2} R. Roth, J. Langhammer, A. Calci, S. Binder, P. Navrátil, Phys. Rev. Lett. 107, 072501 (2011).
\bibitem{srg}
R. Roth, A. Calci, J. Langhammer, S. Binder,
Phys. Rev. C 90, 024325 (2014).
\bibitem{lenpicNN}
E. Epelbaum, H. Krebs, U.-G. Mei\ss{}ner,
Phys. Rev. Lett. 115, 122301 (2015).
\bibitem{lenpic3N}
K. Hebeler, H. Krebs, E. Epelbaum, J. Golak, R. Skibinski,
Phys. Rev. C 91, 044001 (2015).
\bibitem{lenpicfew}
S. Binder \emph{et al.}, Phys. Rev. C 93, 044002 (2016).
\bibitem{sensitivity}
A. Calci, R. Roth,
Phys. Rev. C 94, 014322 (2016).
\end{thebibliographynotitle}

\addcontentsline{toc}{section}{H.~Wita{\l}a et al.: ${\rm Application~of~new~N}^2{\rm LO~3NF's~in~calculations}$\protect\newline ${\rm ~~~~~~~~~~~~~~~~~~~of~3N~reactions}$}
\newabstract 
\begin{center}
{\large\bf Application of new N$^2$LO 3NF's
 in calculations of 3N reactions}\\[0.5cm]
{\bf Henryk Wita{\l}a}$^1$,\let\thefootnote\relax\footnote{This work was supported by the Polish
National Science Center under Grant No. DEC-2013/10/M/ST2/00420.
The numerical calculations have been performed on the
 supercomputer cluster of the JSC, J\"ulich, Germany.} Jacek Golak$^1$, Roman Skibi\'nski$^1$,
and Kacper Topolnicki$^1$  \\[0.3cm]
$^1$M. Smoluchowski Institute of Physics, Jagiellonian
 University,\\
PL-30348 Krak\'ow, Poland\\[0.3cm]
\end{center}

Solving 3N scattering exactly in a numerical sense up to energies below the
pion production threshold allows one
to test  the 3N Hamiltonian based on modern
NN potentials and 3NF's. At higher energies (above $\approx 60$~MeV)
for some observables large
3NF effects are predicted when standard NN interactions
(AV18~\cite{av18}, CDBonn~\cite{cdb}, NijmI and II~\cite{nijm}) are combined
 with  (se\-mi)phe\-no\-me\-no\-lo\-gi\-cal
 models of 3NF's such as TM~\cite{tm3nf} or Urbana IX~\cite{uIX}).
 The large discrepancy between the theory and experimental data
 for the total cross section and
 in the minimum of the elastic scattering
 cross section
 obtained with NN forces
 only, seen for energies above $\approx 60$~MeV, is removed for energies below
$\approx 140$~MeV when 3NF's, which reproduce the experimental
triton binding energy, are included~\cite{wit98},~\cite{wit2001}.
 A similar behavior shows up for the high energy deuteron vector analyzing
power $A_y(d)$~\cite{wit2001}. But there are many spin
observables for which large 3NF effects are predicted and where
 the TM and the Urbana IX 3NF do not reproduce the data~\cite{wit2001}. This is
the case e.g. for the nucleon analyzing power $A_y$~\cite{wit2001}
and  for the deuteron tensor analyzing powers~\cite{wit2001}. In none of these
 cases  the data  can be reproduced by pure 2N force predictions.

The large discrepancies  at higher energies between data and theory in
elastic Nd scattering
  which cannot be removed by taking into account  standard  3NF's require
to study the magnitude of relativistic
effects. They were found to be small for the elastic scattering cross
section and negligible for  spin-observables at higher energies~\cite{wit2007}.

The small size of relativistic effects indicates that very probably the
short range contributions to the 3NF are responsible for the higher energy
elastic scattering discrepancies. The recently constructed
 new generation of chiral
NN potentials up to N$^4$LO with an appropriate regularization in the
 coordinate space~\cite{new1},~\cite{new2} made it possible to  reduce significantly
 finite-cutoff artifacts present when using the
nonlocal momentum-space regulator employed in the chiral NN potentials of
Refs.~\cite{epel2006},~\cite{machrep}.
 Application of these new NN potentials does
 not lead to distortions in the cross
section minimum of the higher energy elastic Nd scattering that were found in
Ref.~\cite{witjour}.

Application of improved chiral NN
interactions  up to  N$^4$LO order of chiral expansion~\cite{binder2016}
 combined with
 N$^2$LO 3NF's   supports conclusions obtained
with standard forces. The observed pattern of higher energy discrepancies
between data and theory resembles that obtained with standard forces.
It can be expected that an application of
consistent chiral NN and 3NF's
up to N$^3$LO will play an important role in understanding of  elastic
scattering and breakup reactions at higher energies.

At low energies effects of a 3NF are rather small  and some serious
discrepancies to data remain even when a 3NF is included.
 The prominent examples are the vector analyzing
power in elastic Nd scattering and
 cross sections for symmetric-space-star (SST) configuration of the
Nd breakup~\cite{glo96}.
They present serious problem for explanation in terms of present day forces.
For SST at $13$~MeV the nd~\cite{sst} and pd~\cite{koln13} breakup data
clearly differ and
 are far away from theory. The
calculations of the pd breakup with inclusion of the pp Coulomb
force \cite{deltuvabr}
revealed only very small Coulomb force effects for this configuration.
Since at that energy the SST configuration is practically dominated by S-wave
NN force components,  the big difference between pd and nd data could suggest
  large charge-symmetry breaking in the $^1S_0$ partial wave. On the other hand
 the discrepancy to theory would imply that our knowledge of the
$^1S_0$ pp and nn low
energy forces is probably insufficient.

\setlength{\bibsep}{0.0em}
\begin{thebibliographynotitle}{99}
\bibitem{av18} R.~B.~Wiringa, V.~G.~J.~Stoks, R.~Schiavilla,
 Phys. Rev. C{\bf 51}, 38 (1995).
\bibitem{cdb} R.~Machleidt,
 Phys. Rev. C{\bf 63}, 024001 (2001).
\bibitem{nijm} V.~G.~J.~Stoks et al.,
 Phys. Rev. C{\bf 49}, 2950 (1994).
\bibitem{tm3nf} S.~A.~Coon et al.,
 Nucl. Phys. {\bf A317}, 242 (1979).
\bibitem{uIX} B.~S.~Pudliner et al.,
 Phys. Rev. C{\bf 56}, 1720 (1997).
\bibitem{wit98} H.~Wita{\l}a et al.,
  Phys. Rev. Lett. {\bf 81}, 1183 (1998).
\bibitem{wit2001} H.~Wita{\l}a et al.,
  Phys. Rev. C{\bf 63}, 024007 (2001).
\bibitem{wit2007} H.~Wita{\l}a {\it et al.}:
Phys. Rev. C{\bf{ 77}} (2008) 034004.
\bibitem{new1} E. Epelbaum, H. Krebs and U.-G. Mei{\ss}ner, Eur. Phys.
J. A {\bf 51}, no. 5, 53 (2015).
\bibitem{new2}  E. Epelbaum, H. Krebs and U.-G. Mei{\ss}ner,
Phys. Rev. Lett. {\bf 115}, 122301 (2015).
\bibitem{epel2006} E.~Epelbaum,
 Prog. Part. Nucl. Phys.  {\bf 57}, 654  (2006)
\bibitem{machrep} R.~Machleidt and D.R.~Entem,
  Phys. Rep.{\bf 503}, 1 (2011).
\bibitem{witjour} H.~Wita{\l}a  et al.,   J. Phys. G{\bf{ 41}}, 094011
(2014).
\bibitem{binder2016} S.~Binder {\it et al.}:
Phys. Rev. C {\bf 93} (2016) 044002.
\bibitem{glo96} W.~Gl\"ockle et al.,
  Phys. Rep.{\bf 274}, 107 (1996).
\bibitem{sst} H. R. Setze et al.,
  Phys. Lett. {\bf{B388}}, 229  (1996).
\bibitem{koln13} G. Rauprich et al.
 Nucl. Phys. \textbf{A535}, 313  (1991).
\bibitem{deltuvabr} A. Deltuva, A.C. Fonseca, and P.U. Sauer,
 Phys. Rev. C{\bf 72}, 054004 (2005).
\end{thebibliographynotitle}

\addcontentsline{toc}{section}{A.~Calci: ${\rm Probing~and~Constraining~chiral~Interactions~in~Nuclear}$ \protect\newline 
                             ${\rm ~~~~~~~~~~Structure~and~Reaction~Calculations}$}

\newabstract 
\begin{center}
{\large\bf Probing and Constraining chiral
Interactions \\ in Nuclear Structure
and Reaction Calculations}\\[0.5cm]
{\bf Angelo Calci}$^1$  \\[0.3cm]
$^1$TRIUMF, 4004 Wesbrook Mall, Vancouver, British Columbia, V6T 2A3, Canada\\[0.3cm]
\end{center}

In view of the current developments to derive nuclear interactions from chiral effective field theory~\cite{EnMa03},\cite{Navr07},\cite{EpGl04},\cite{EpGl05},\cite{EpKr15},\cite{EpKr15b},\cite{EkBa13},\cite{EkJa15} the improvements of the No-Core Shell Model (NCSM) based ab initio approaches are of particular importance.
Concepts such as using Importance-Truncated (IT) model spaces applied in the IT-NCSM~\cite{Roth09},\cite{RoNa07} allow us to study spectroscopy of bound-state systems throughout the p- and lower sd-shell within controlled approximations. In particular, the consideration of cluster formations in certain nuclei that is exploited in the NCSM with Continuum (NCSMC)~\cite{BaNa13b},\cite{BaNa13c},\cite{NaQu16} enables us to study weakly-bound systems, resonances and nuclear reactions on the same footing. The nuclear systems that can be computed with these ab initio methods often show a strong sensitivity to the details of the nuclear interactions and thus can be used to probe and constrain them.
By introducing modifications of the chiral 3N interactions and studying chiral interactions that differ in the chiral order truncation, the regularization scheme and the fit procedure of the chiral constants~\cite{EnMa03},\cite{Navr07},\cite{EpGl04},\cite{EpGl05},
\cite{EpKr15},\cite{EpKr15b},\cite{EkBa13},\cite{EkJa15} we can perform a first step towards an uncertainty analysis in nuclear spectroscopy and moreover identify a strong correlated sensitivity of observables such as the excitation energy of the first $1^+$ state in ${}^{10}B$ and ${}^{12}C$~\cite{CaRo16}.

An accurate description of the low-lying spectrum in the ${}^{11}Be$ system that is dominated by an $n+{}^{10}Be$ halo structure has been proven to be extremely challenging~\cite{CaNa16}. With state of the art ab initio NCSMC calculations we demonstrate that the reproduction of the parity inversion for the two weakly-bound states depend on the details of the chiral NN and 3N interactions. The best reproduction of the energy spectrum and E1 transitions in ${}^{11}Be$ including the parity-inversion properties is achieved with the N${}^{2}$LO${}_{\text{SAT}}$ interaction~\cite{EkJa15} that generally shows a good reproduction of long-range properties such as radii for this mass regime.

Due to the inclusion of explicit 3N interactions, the computational costs of the NCSMC calculations limits the applications range and number of interactions that can be investigated.
However, by combining the concept of the NCSMC and the multi-reference normal-ordering (MR-NO) approach~\cite{GeCa16} allows us to include the 3N interactions in an approximative manner that is extremely accurate and reduces the computational costs by about two orders of magnitude.
This new development allows us to increase the mass number and study as a first example the ${}^{12}N$ system that is dominated by the $p+{}^{11}C$ structure, but also revisit lighter reactions such as $n+{}^{4}He$ for the large number of novel chiral potentials.
The splitting of the {\it{P}}$_{3/2}$ and {\it{P}}$_{1/2}$ $n+{}^{4}He$ phase shifts shows a strong sensitivity to the 3N force and constitutes an ideal candidate to constrain future chiral interactions.

\setlength{\bibsep}{0.0em}
\begin{thebibliographynotitle}{99}

\bibitem{EnMa03}
D.~R. Entem and R.~Machleidt, {\em Phys. Rev.
  C}, vol.~68, p.~041001(R), 2003.

\bibitem{Navr07}
P.~Navr{\'a}til,  {\em Few-Body Syst.}, vol.~41, pp.~117--140, 2007.

\bibitem{EpGl04}
E.~Epelbaum, W.~Gl\"ockle, and U.-G. Mei\ss{}ner,  {\em Eur. Phys. J. A}, vol.~19, p.~401, 2004.

\bibitem{EpGl05}
E.~Epelbaum, W.~Gl\"ockle, and U.-G. Mei\ss{}ner,  {\em Nucl. Phys. A}, vol.~747,
  p.~362, 2005.

\bibitem{EpKr15}
E.~Epelbaum, H.~Krebs, and U.-G. Mei\ss{}ner,  {\em Phys. Rev. Lett.},
  vol.~115, p.~122301,  2015.

\bibitem{EpKr15b}
E.~Epelbaum, H.~Krebs, and U.-G. Mei{\ss}ner,  {\em Eur. Phys. J.
  A}, vol.~51, no.~5, 2015.

\bibitem{EkBa13}
A.~Ekstr\"om, G.~Baardsen, C.~Forss\'en, G.~Hagen, M.~Hjorth-Jensen, G.~R.
  Jansen, R.~Machleidt, W.~Nazarewicz, T.~Papenbrock, J.~Sarich, and S.~M.
  Wild,  {\em Phys. Rev. Lett.}, vol.~110, p.~192502,
   2013.

\bibitem{EkJa15}
A.~Ekstr\"om, G.~R. Jansen, K.~A. Wendt, G.~Hagen, T.~Papenbrock, B.~D.
  Carlsson, C.~Forss\'en, M.~Hjorth-Jensen, P.~Navr\'atil, and W.~Nazarewicz,
  {\em Phys. Rev. C}, vol.~91, p.~051301(R),  2015.

\bibitem{Roth09}
R.~Roth,  {\em Phys. Rev. C}, vol.~79, p.~064324, 2009.

\bibitem{RoNa07}
R.~Roth and P.~Navr\'atil, {\em Phys. Rev.
  Lett.}, vol.~99, p.~092501, 2007.

\bibitem{BaNa13b}
S.~Baroni, P.~Navr\'atil, and S.~Quaglioni,  {\em Phys. Rev. Lett.}, vol.~110,
  p.~022505,  2013.

\bibitem{BaNa13c}
S.~Baroni, P.~Navr\'atil, and S.~Quaglioni,  {\em Phys. Rev. C}, vol.~87, p.~034326,   2013.

\bibitem{NaQu16}
P.~Navr\'atil, S.~Quaglioni, G.~Hupin, C.~Romero-Redondo, and A.~Calci,
 {\em  Phys. Scr.}, vol.~91, no.~5, p.~053002, 2016.

\bibitem{CaRo16}
A.~Calci and R.~Roth,  {\em Phys. Rev. C}, vol.~94,
  p.~014322,  2016.

\bibitem{CaNa16}
A.~Calci, P.~Navr\'atil, R.~Roth, J.~Dohet-Eraly, S.~Quaglioni, and G.~Hupin,
 {\em Phys. Rev. Lett.}, vol.~117, p.~242501,  2016.

\bibitem{GeCa16}
E.~Gebrerufael, A.~Calci, and R.~Roth,  {\em Phys. Rev. C},
  vol.~93, p.~031301,  2016.

\end{thebibliographynotitle}

\addcontentsline{toc}{section}{P.~Reinert et al.: ${\rm Partial{-}Wave~Analysis~of~NN~scattering~data}$ \protect\newline 
                             ${\rm ~~~~~~~~~~~~~~~~~~~~at~fifth~order~in~chiral~EFT}$}
\newabstract 
\begin{center}
{\large\bf Partial-Wave Analysis of NN scattering data at fifth order in chiral EFT}\\[0.5cm]
{\bf Patrick Reinert}$^1$, Evgeny Epelbaum$^1$, and Hermann Krebs$^1$  \\[0.3cm]
$^1$Institut f\"ur Theoretische Physik II, Ruhr-Universit\"at Bochum,\\
44780 Bochum, Germany\\[0.3cm]
\end{center}
In effective field theories, free parameters known as low-energy constants (LECs) emerge, which are not constrained by symmetry and instead have to be fixed by experimental data. In the case of our improved chiral NN potential \cite{NNPotentialN4LO},~\cite{NNPotentialN3LO}, the nucleon-nucleon (NN) contact LECs have (up to now) been determined by a phase shift fit to the Nijmegen partial-wave analysis \cite{NijmegenPWA}. To overcome any possible model dependence of the Nijmegen analysis and to account for scattering data published after it, we determine the LECs here directly from experimental neutron-proton and proton-proton scattering data. We employ the same treatment of electromagnetic effects as in \cite{NijmegenPWA}.

The experimental data are taken from the 2013 Granada database \cite{Granada2013} of mutually compatible scattering data. In order to extend the energy range of the fit to $T_{lab}=0-300$ MeV and to account for the decreasing accuracy of the chiral expansion at higher energies, we estimate the theoretical uncertainties of the scattering observables, as detailed e.g. in \cite{NNPotentialN4LO}, and combine them with the experimental errors.

A small set of high-precision proton-proton data at $T_{lab} > 140$ MeV can not be described to its experimental precision by the N$^4$LO predictions (although the deviation lies within the theoretical uncertainty). In order to probe the sensitivity of these observables to partial waves which are not parametrized at N$^4$LO, we introduce a N$^4$LO$^+$ potential, which differs from the one of \cite{NNPotentialN4LO} by the inclusion of N$^5$LO contact terms in F-waves. We find that the description of the aforementioned outliers is significantly improved and we are now able to describe, after fitting, all proton-proton scattering data in the range of 0-300 MeV with a $\chi^2$/datum of 1.04, while for neutron-proton data we have a $\chi^2$/datum of 1.11.

The obtained phase shifts are in good agreement with the Nijmegen and Granada analyses and the description of experimental data at N$^4$LO$^+$ is comparable to the one of high-quality phenomenological NN potentials, in terms of precision.

\setlength{\bibsep}{0.0em}
\begin{thebibliographynotitle}{99}
\bibitem{NNPotentialN4LO}
E.~Epelbaum, H.~Krebs and U.G.~Mei{\ss}ner, Phys. Rev. Lett. 115, 122301 (2015)
\bibitem{NNPotentialN3LO}
E.~Epelbaum, H.~Krebs and U.G.~Mei{\ss}ner, Eur. Phys. J. A 51, 53 (2015)
\bibitem{NijmegenPWA}
V.~G.~J.~Stoks, R.~A.~M.~Klomp, M.~C.~M.~Rentmeester and J.~J.~de Swart, Phys. Rev. C 48, 792 (1993).
\bibitem{Granada2013}
R.~Navarro Pérez, J.~E.~Amaro and E.~Ruiz Arriola, Phys. Rev. C 88, 064002 (2013)
\end{thebibliographynotitle}

\addcontentsline{toc}{section}{J.~P.~Vary: ${\rm No{-}Core~Shell~Model:~Quantifying~the~Observables'}$ \protect\newline 
                ${\rm ~~~~~~~~~~~~~Uncertainties}$}
\newabstract 
\begin{center}
{\large\bf {\it \bf{Ab Initio}} No-Core Shell Model: \\ Quantifying the Observables' Uncertainties}\\[0.5cm]
{\bf James P. Vary}$^1$  \\[0.3cm]
$^1$Department of Physics and Astronomy, Iowa State University\\
Ames, Iowa   50011, USA \\[0.3cm]
\end{center}

Since our goal is to preserve predictive power with quantified uncertainties, it is important to
have a solid grasp of all sources of theoretical and numerical uncertainty.
I survey uncertainties that appear in current applications of the
{\it ab initio} No-Core Shell Model (NCSM)~\cite{Navratil:2000ww},\cite{Navratil:2000gs},\cite{Barrett:2013nh}
with Hamiltonians developed in
Chiral Effective Field Theory ($\chi$EFT)~\cite{Epelbaum},\cite{Entem:2003ft},\cite{Machleidt:2011zz},
\cite{Epelbaum:2014sza}.

Under the banner of uncertainties
associated with $\chi$EFT, I include several topics and comments:

\begin{itemize}
\item{Fitting of LECs, NN data error propagation~\cite{Perez:2015bqa}  (other LENPIC teams)}
\item{Choice of regulator (results presented here are for R = 1.0 fm so far)}
\item{Truncation at a fixed Chiral order (presenting here a revised method to estimate the uncertainty for the ground
state energies of nuclei) }
\item{Numerical uncertainty at fixed [$N_{\rm max}$, $\hbar\Omega$] (~1 keV in total gs energy)}
\item{Extrapolation uncertainty (new results here for gs energies)~\cite{Coon:2012ab},\cite{Furnstahl:2012qg}}
\end{itemize}

Under the banner of uncertainties associated with the NCSM, I also include several topics and comments:

\begin{itemize}
\item{Truncation vs OLS renormalization applied to the effective operators (large effects shown here)}
\item{Rank of OLS-derived operator truncation (2-body, 3-body, . . . )}
\item{Other approximations, if adopted, such as normal ordering approximation, importance truncation, . . . }
\end{itemize}

The main conclusions are twofold:

\begin{itemize}
\item{Uncertainty versus chiral order appears consistent when adopting an average relative momentum
scale from the ground state total relative kinetic energy to define a nucleus-dependent chiral expansion parameter Q.}
\item{OLS succeeds in renormalizing the IR and UV scales in initial applications to electroweak operators.}
\end{itemize}

The outlook for these lines of investigation are promising and include:
\begin{itemize}
\item{Novel approach to scattering is now established and used to predict the tetraneutron.
This opens a path for scattering applications with chiral interactions
in light nuclei~\cite{Shirokov:2016ywq},\cite{Shirokov:2016thl}.}
\item{Major additional efforts are needed to develop and apply these methods:
effective Hamiltonians, effective electroweak operators, many-body methods, computational algorithms, . . . }
\end{itemize}

I am especially indebted to Pieter Maris for many stimulating discussions of uncertainty quantification. I thank my students for diligent efforts to compute some of the effective operator results shown here. I thank the Cracow group for their warm hospitality and for their support of this meeting. This work is sponsored by th US NSF grant number PHY1516181 and by the US DOE DE-FG02-87ER40371.

\setlength{\bibsep}{0.0em}
\begin{thebibliographynotitle}{99}
\bibitem{Navratil:2000ww}
  P.~Navratil, J.~P.~Vary and B.~R.~Barrett,
  Phys.\ Rev.\ Lett.\  {\bf 84}, 5728 (2000)
  [nucl-th/0004058].

\bibitem{Navratil:2000gs}
  P.~Navratil, J.~P.~Vary and B.~R.~Barrett,
  Phys.\ Rev.\ C {\bf 62}, 054311 (2000).

\bibitem{Barrett:2013nh}
  B.~R.~Barrett, P.~Navratil and J.~P.~Vary,
  Prog.\ Part.\ Nucl.\ Phys.\  {\bf 69}, 131 (2013).

\bibitem{Epelbaum} E. ~Epelbaum, W. ~Gl\"ockle, and Ulf-G. ~Mei{\ss}ner,
Nucl. Phys. A {\bf 637}, 107 (1998); {\bf 671}, 295 (2000).

\bibitem{Entem:2003ft}
  D.~R.~Entem and R.~Machleidt,
  Phys.\ Rev.\ C {\bf 68} (2003) 041001
  [nucl-th/0304018].

\bibitem{Machleidt:2011zz}
  R.~Machleidt and D.~R.~Entem,
  Phys.\ Rept.\  {\bf 503}, 1 (2011)
  [arXiv:1105.2919 [nucl-th]].

\bibitem{Epelbaum:2014sza}
  E.~Epelbaum, H.~Krebs and U.~G.~Mei{\ss}ner,
  Phys.\ Rev.\ Lett.\  {\bf 115}, no. 12, 122301 (2015)
  [arXiv:1412.4623 [nucl-th]].

\bibitem{Perez:2015bqa}
R.~Navarro P<8e>rez, J.~E.~Amaro, E.~Ruiz Arriola, P.~Maris and J.~P.~Vary,
  Phys.\ Rev.\ C {\bf 92}, no. 6, 064003 (2015)
  [arXiv:1510.02544 [nucl-th]].

\bibitem{Coon:2012ab}
  S.~A.~Coon, M.~I.~Avetian, M.~K.~G.~Kruse, U.~van Kolck, P.~Maris and J.~P.~Vary,
  Phys.\ Rev.\ C {\bf 86}, 054002 (2012)
  [arXiv:1205.3230 [nucl-th]].

\bibitem{Furnstahl:2012qg}
  R.~J.~Furnstahl, G.~Hagen and T.~Papenbrock,
  Phys.\ Rev.\ C {\bf 86}, 031301 (2012)
  [arXiv:1207.6100 [nucl-th]].

\bibitem{Shirokov:2016ywq}
  A.~M.~Shirokov, G.~Papadimitriou, A.~I.~Mazur, I.~A.~Mazur, R.~Roth and J.~P.~Vary,
  Phys.\ Rev.\ Lett.\  {\bf 117} (2016) 182502
  [arXiv:1607.05631 [nucl-th]].

\bibitem{Shirokov:2016thl}
  A.~M.~Shirokov, A.~I.~Mazur, I.~A.~Mazur and J.~P.~Vary,
  Phys.\ Rev.\ C {\bf 94}, no. 6, 064320 (2016)
  [arXiv:1608.05885 [nucl-th]].

\end{thebibliographynotitle}

\addcontentsline{toc}{section}{H.~Kamada et al.: ${\rm  Relativistic~Faddeev~Calculations}$}
\newabstract 
\begin{center}
{\large\bf Relativistic Faddeev Calculations}\\[0.5cm]
{\bf Hiroyuki Kamada}$^1$, Henryk Wita\l a$^2$, Jacek Golak$^2$, Roman Skibi\'nski$^2$, \\
Oleksandr Shebeko$^3$, and Adam Arslanaliev$^4$\\[0.3cm]
$^1$Department of Physics, Faculty of Engineering, Kyushu Institute of Technology, 1-1 Sensuicho Tobata,
Kitakyushu 804-8550, Japan \\[0.3cm]
$^2$M. Smoluchowski Institute of Physics, Jagiellonian University, 30048 Krak\'ow, Poland
\\[0.3cm]
$^3$NSC Kharkov Institute of Physics and Technology, NAS of Ukraine, Kharkiv, Ukraine \\[0.3cm]
$^4$
The Karazin University, Kharkiv, Ukraine
\end{center}

Gl\"ockle {\it et al.} started to study \cite{Gloeckle86} relativity in the three-nucleon (3N)
system under the Bakamjian-Thomas formalism \cite{Bakamjian}, which
belongs to the relativistic quantum mechanics and is dictated by the  Poincar\'e algebra.
Since a relativistic two-nucleon (2N) potential is not easily provided, we need some schemes
 which would allow us to transform
a nonrelativistic potential into the corresponding relativistic one.
Such schemes are required to fulfill the condition that the generated
 relativistic potential
yields the same  observables in the 2N system as the original
 nonrelativistic potential.
There are two schemes which satisfy that condition. One was
 proposed by Coester {\it et al.}
\cite{Coester} and we call it the CPS scheme. It requires a solution
of a nonlinear integral equation, which can be achieved numerically by an iteration
method \cite{Kamada07}. The other scheme is a momentum scaling method (MSM)  \cite{Kamada98},
realized by an elaborate change of momentum variables.
The above-mentioned schemes are examples of simple transformations, while we are
actually interested in a comparison
between the original nonrelativistic and the modified relativistic
potential predictions in the 3N system.
 In the case of the triton binding energy
we have already shown this comparison \cite{Kamada10}, demonstrating that the
difference is smaller when the CPS scheme is employed.

On the other hand, the Kharkov model \cite{Dubovyk2010} provides directly the relativistic 2N potential
so no transformation scheme is needed in this case.

At this workshop we present the relativistic results of the triton binding energies not only for the Kharkov potential \footnote{Only the 5channel result of the Kharkov potential was already shown in \cite{Kamada17}.
}
but also for the new N4LO chiral potential \cite{N4LO} and, additionally, for the older realistic CDBonn potential \cite{CDBonn}.
In Table \ref{tab:1} the triton binding energies for these potentials,
are demonstrated.

\begin{table}
\caption{The theoretical predictions for the triton binding energy (in MeV), resulting from the solutions of
the relativistic and nonrelativistic Faddeev equations with 42 3N-partial-wave states
 ($j_{max}=5$).
The numbers in brackets are obtained by the CPS scheme~\cite{Coester},
used to transform the nonrelativistic potentials into the relativistic ones,
and for the opposite transition in the case of
the Kharkov potential. 
}
\label{tab:1}       
\begin{center}
\begin{tabular}{llll}
\hline\noalign{\smallskip}
   Potential type       &  Nonrelativistic calc.  &  Relativistic calc. &  Difference \\
\noalign{\smallskip}\hline\noalign{\smallskip}
CDBonn \cite{CDBonn}              & ~~-8.249  & (~-8.150 )   &  0.099   \\
N4LO (R=0.9~fm)          & ~~-7.832  & (~-7.706 )   &  0.126    \\
N4LO (R=1.0~fm)          & ~~-7.867  & (~-7.748 )   & 0.119   \\
N4LO (R=1.1~fm)          & ~~-7.847  & (~-7.733 ) & 0.115 \\
Kharkov \cite{Dubovyk2010}                 & (~-7.528 ) & ~~-7.641  &   0.067  \\
\noalign{\smallskip}\hline
\end{tabular}
\end{center}
\end{table}

Using the CDBonn  potential  and the Tucson-Melbourne 3N force we have investigated the Nd elastic scattering \cite{Witala05}.
Relativistic calculations ~\cite{Witala05},\cite{Witala11},\cite{Sekiguchi2005} show
 only small effects for elastic scattering cross sections and practically
no effects for spin observables.
Relativistic effects in the nucleon-induced deuteron breakup have been
 investigated in~\cite{Witala2006}.
The study of Nd scattering based on the Kharkov potential is also in progress.

This work was partially supported by the Polish National Science
Center under Grants No. DEC-2013/10/M/ST2/00420, and by Grant-in-Aid for Scientific Research (B)
No: 16H04377, Japan Society for the Promotion of Science (JSPS).
The numerical calculations were partially performed on the interactive server
at RCNP, Osaka University, Japan, and in
the JSC, J\"ulich, Germany.

\vspace{-6mm}

\setlength{\bibsep}{0.0em}
\begin{thebibliographynotitle}{99}

\bibitem{Gloeckle86} W. Gl\"ockle, T-S. H. Lee, and  F. Coester,
Phys. Rev. C {\bf 33}, (1986) 709.
\bibitem{Bakamjian} B. Bakamjian, L.H. Thomas, Phys. Rev. {\bf 92}, (1953) 1300.

\bibitem{Coester} F. Coester, Steven C. Pieper,  F.J.D. Serduke,
Phys. Rev. C {\bf 11}, (1975) 1.

\bibitem{Kamada07} H. Kamada, W. Gl\"ockle,
 Phys. Lett. B{\bf 655}, (2007) 119.

\bibitem{Kamada98} H. Kamada, W. Gl\"ockle,
 Phys. Rev. Lett. {\bf 80}, (1998) 2457.

 \bibitem{Kamada10} H. Kamada, {\it et al.},
EPJ Web of Conf. {\bf 3}, (2010) 05025.

\bibitem{Dubovyk2010}  I. Dubovyk,  O. Shebeko,
 Few-Body Syst. {\bf 48}, (2010) 109.

\bibitem{N4LO} E. Epelbaum {\it et al.}, Eur. Phys. J. {\bf A51} (2015) 53.

\bibitem{CDBonn} R. Machleidt, F. Sammarruca, Y. Song,
 Phys. Rev. C {\bf 53}, (1996) R1483.

\bibitem{Witala05}  H. Wita\l a {\it et al.},
 Phys. Rev. C {\bf 71}, (2005) 054001.

\bibitem{Witala11}  H. Wita\l a, et al., 
Phys. Rev.  C {\bf 83}, (2011) 044001 ; Erratum, Rev.  C {\bf 88},(2013)  069904(E).

\bibitem{Sekiguchi2005}  K. Sekiguchi {\it et al.},
Phys. Rev. Lett. {\bf 95}, (2005) 162301;
Y. Maeda {\it et al.}, Phys. Rev. C {\bf 76}, (2007) 014004.

\bibitem{Kamada17} H.Kamada, O.Shebeko, and A.Arslanaliev, Few-Body Syst. {\bf 58}, (2017) 70.

\bibitem{Witala2006}  H. Wita\l a, J. Golak, R. Skibi\'nski,
Phys. Lett. B {\bf 634}, 374 (2006)

\end{thebibliographynotitle}

\begin{figure}[!hb]
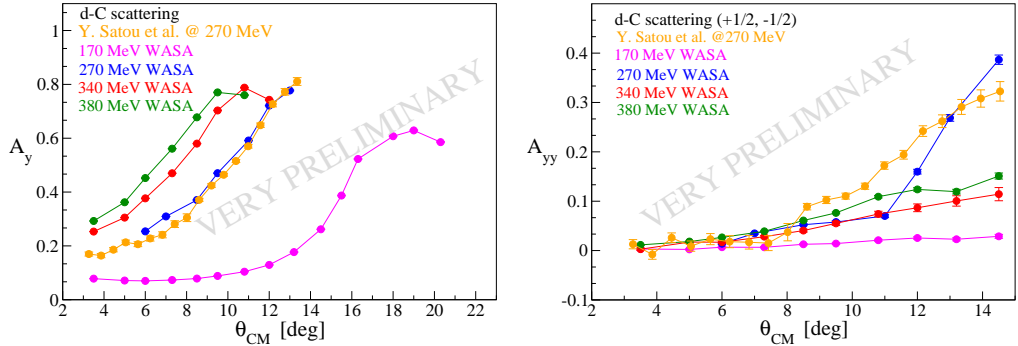

   \begin{minipage}{.5\textwidth}
      \hspace{3.0mm}
  \includegraphics[width=.9\textwidth]{vect_dC.eps}
   \end{minipage}
    \begin{minipage}{0.5\textwidth}
   \includegraphics[width=.9\textwidth]{tens_dC.eps}
  \end{minipage}
\caption{\small {Measurement of the deuteron elastic scattering $A_{y}$ and $A_{yy}$ analyzing powers at energies listed in the panel.
 The data at 270 MeV (orange) were used to calibrate the polarization values. The data are very preliminary. }}
\label{fig1}
\end{figure}

\addcontentsline{toc}{section}{I.~Ciepa{\l} et al.: ${\rm Recent~measurement~of~}^{12}{\rm C(d,d)~tensor~and~vector}$ \protect\newline 
${\rm ~~~~~~~~~~~~~~~~~~analyzing~powers~at~the~energy~range~170{-}380~MeV~}$ \protect\newline 
${\rm ~~~~~~~~~~~~~~~~~~in~a~view~of~the~charged~particles~EDM~studies}$}
\newabstract 
\begin{center}
{\large\bf Recent measurement of $^{12}$C(d, d) tensor and vector analyzing powers
at the energy range 170-380 MeV in a view of the charged particles EDM studies}\\[0.5cm]
{\bf Izabela Ciepa\l{}$^1$} and Ed Stephenson$^2$  for the JEDI Collaboration  \\[0.3cm]
$^1$Institute of Nuclear Physics Polish Academy of Sciences, PL-31342 Krak\`{o}w, Poland,\\[0.3cm]
$^2$Indiana University Center for Spacetime Symmetries, Department of Physics
107 S. Indiana Avenue, Bloomington, Indiana 47405 USA\\[0.3cm]

\end{center}
Permanent  electric dipole moments (EDM) of fundamental  particles  violate  both  time invariance T and  parity P.
Assuming  the CPT-theorem, this leads to CP violation, which is needed to explain
the matter over antimatter dominance in the universe.
Standard Model predictions for this EDM lead to extremely small value - typically many orders of magnitude below
current experimental limits. However, extensions of the Standard Model, e.g. Supersymmetric Models,
predict much larger EDM. The observation of non zero EDM would represent
a clear sign of new physics beyond the Standard Model.
Researchers  have  been  searching  for  EDM  in  neutral  particles,  especially  neutrons,  for  more
than 50 years,  by trapping and cooling particles in small volumes and using strong electric fields.
Despite  an  enormous  improvement  in  sensitivity, these  experiments  have  only  produced
upper bound which currently is  10$^{-26}$ e$\cdot$cm for the neutron.
Theoretical models predict many CP-violating mechanisms and therefore, not only the neutron EDM
measurement is important, but so are other hadrons and (light) nuclei,
e.g.\ the proton, deuteron, ${}^{3}$He, etc. This will help  to isolate the specific
CP-violating source(s). Recently, the two-nucleon contributions to the deuteron EDM
were calculated in the framework of effective field theory up-to-and-including N2LO~\cite{Bsa}. \\
\indent The JEDI (J\"{u}lich Electric Dipole moment Investigations) collaboration \cite{jedi} has started investigations of a direct EDM
measurement of charged hadrons at a storage ring \cite{Fat},~\cite{Gui} .
The basic idea is to align the beam polarization along the momentum, and keep the beam
circulating while interacting with the radial electric field always present in the particle frame. The
EDM signal would then be detected as tiny changes (at the level of a microradian) in the vertical polarization during
its storage time. To measure such small effects one needs to build a dedicated high precision polarimeter.
To reach a required sensitivity of 10$^{-29}$ e$\cdot$cm, the polarimeter system must fulfill the following requirements:
very stable long-term (long beam storage time), efficient (up to 1\%), sensitive to the vertical and
horizontal polarization components (high analyzing powers of at least 0.5) and  strong radiation hardness.
To meet to above expectations in a first step the $^{12}$C($\vec{d}$, d) tensor and vector analyzing powers
in the energy range 170-380 MeV were measured with the use of the WASA Forward detector at COSY \cite{Bar}. These data will be used to produce
realistic Monte Carlo simulations of detector responses for the polarimeter design. In parallel, the prototype of the polarimeter
is under the development at COSY.\\
\indent Two pure vector states (-$\frac{2}{3}$, 0), (+$\frac{2}{3}$, 0) and two mixed states (-1, +1), \\(+$\frac{1}{2}$, -$\frac{1}{2}$)
were used in the experiment.
The vector and tensor polarizations were first calibrated using the data at 270 MeV \cite{Sat}.
For the vector state the polarization was about 85\% of maximum and for the tensor 50\% and 83\%, respectively.
Using the so-called ``cross ratio'' \cite{Ohl}, which nominally cancels the systematic errors up to second order \cite{Bra},
$A_{y}$ and $A_{yy}$ were calculated. In Fig. \ref{fig1} the preliminary data obtained are presented,
together with the data used for the calibration \cite{Sat}.  The vector and tensor analyzing powers were
obtained at different energies of 270 MeV and above. In a peak $A_{y}$ achieves value of about 0.8 and remains the same for all energies.
In the case of $A_{yy}$ the sensitivity is less.

\vspace{5mm}

\setlength{\bibsep}{0.0em}
\begin{thebibliographynotitle}{99}
\bibitem{Bsa}J. Bsaisou et al., Annals of Physics 359 (2015) 317.
\bibitem{jedi}http://collaborations.fz-juelich.de/ikp/jedi/
\bibitem{Fat}F.J.M. Farley et al., Phys. Rev. Lett. 93 (2004) 052001.
\bibitem{Gui}G. Guidoboni, E.J. Stephenson, et al., Phys. Rev. Lett. 117 (2016) 054801.
\bibitem{Bar}Chr. Bargholtz et al., Nucl. Instrum. Methods A594 (2008) 339.
\bibitem{Sat}Y. Satou et al., Phys. Lett. B549 (2002) 307.
\bibitem{Ohl}G.G. Ohlsen, P.W. Keaton Jr., Nucl. Instrum. Methods A109 (1973) 41.
\bibitem{Bra}N.P.M. Brantjes, P.W. Keaton Jr., Nucl. Instrum. Methods A664 (2012) 49.

\end{thebibliographynotitle}

\addcontentsline{toc}{section}{H.~Krebs: ${\rm Nuclear~currents~in~chiral~effective~field~theory}$}
\newabstract 
\begin{center}
{\large\bf Nuclear currents in chiral effective field theory}\\[0.5cm]
{\bf Hermann Krebs}  \\[0.3cm]
Institut f\"ur Theoretische Physik II, Ruhr-Universit\"at Bochum,\\
44780 Bochum, Germany\\[0.3cm]
\end{center}

Chiral effective field theory (EFT) is a powerful tool for
description of low energy nuclear phenomena. Based on the symmetries
of Quantum Chromodynamics (QCD) it provides a systematic expansion of
the nuclear forces in small momenta and masses scale.

Within chiral EFT nuclear vector and axial vector currents have been analyzed by different
groups up to the order $Q$ in the chiral expansion (N$^3$LO). The currents have been calculated by using
time-ordered perturbation theory (TOPT, see \cite{Riska:2016cud} and
references therein) and a method of unitary
transformations (UT, see~\cite{Krebs:2016rqz} and references therein). In our recent work within UT~\cite{Krebs:2016rqz} we analyzed the
axial vector current and compared our results with TOPT results. We
found discrepancies for two-pion exchange contributions. It remains to
be seen if there exists a unitary transformation which would bridge
TOPT and UT results.

Within UT method we found a large unitary ambiguity for the axial
vector current which is mainly caused by pion-pole
contributions. These contributions are coming from pion-production
substructures. If we require pion-production substructures of the
axial vector current and the three-nucleon forces to be the same (we call
this requirement matching condition) and
additionally require the currents to be perturbatively renormalizable
we arrive at a unique result. All unitary phases become fixed by
these two conditions.

In practical implementation the currents need to be
regularized. Their regularization, however, should be chosen consistently with
the nuclear forces. Matching condition suggests
the same form of the regulator as in the nuclear forces. The chiral
symmetry, on the other hand, provides an additional constraint which is the continuity
equation. We propose for future investigations to force the continuity equation to be exactly
(not only up to higher order terms) fulfilled for regularized currents
(Siegert approach). Only in this case the cutoff dependence
of the low
energy constant $C_D$  in the axial vector
current and the thee-nucleon forces will be the same. Numerical
implementation within the proposed line are underway.

\setlength{\bibsep}{0.0em}
\begin{thebibliographynotitle}{99}
\bibitem{Riska:2016cud}
  D.~O.~Riska and R.~Schiavilla,
  Int.\ J.\ Mod.\ Phys.\ E {\bf 26}, no. 01n02, 1740022 (2017)
  doi:10.1142/S0218301317400225
  [arXiv:1603.01253 [nucl-th]].
\bibitem{Krebs:2016rqz}
  H.~Krebs, E.~Epelbaum and U.-G.~Mei{\ss}ner,
  Annals Phys.\  {\bf 378}, 317 (2017)
  doi:10.1016/j.aop.2017.01.021
  [arXiv:1610.03569 [nucl-th]].
\end{thebibliographynotitle}

\addcontentsline{toc}{section}{H.~Le et al.: ${\rm Jacobi~No{-}Core~Shell~Model~for~Hypernuclei}$}
\newabstract 
\begin{center}
{\large\bf Jacobi No-core Shell Model for Hypernuclei}\\[0.5cm]
{\bf Hoai Le}$^1$, Ulf-G Mei{\ss}ner$^{1,2}$, and Andreas Nogga$^{1,3}$  \\[0.3cm]
$^1$Institut f\"ur Kernphysik, Institute for Advanced Simulation and J\"ulich Center
for Hadron Physics, Forschungszentrum J\"ulich, D-52425 J\"ulich, Germany \\ [0.3cm]
$^2$Helmholtz-Institut f\"ur Strahlen- und Kernphysik  and Bethe Center
for Theoretical Physics, Universit\"at Bonn, D-53115 Bonn, Germany. \\ [0.3cm]
$^3$Department of Physics and Astronomy, Ohio University, Athens, OH 45701, USA \\[0.3cm]
\end{center}

Hypernuclei have recently attracted a lot of attention due to
new experimental data. Ongoing experiments at international facilities
like JPARC, JLAB, Mainz and FAIR are expected to provide more accurate information on
hypernuclear structure. To establish a direct link between these data and properties of
the hyperon-nucleon (YN) interaction, theoretical predictions for these hypernuclei based on
realistic YN interaction models are vital.

In this talk, we presented our recent work on no-core shell model for light
hypernuclei using a harmonic oscillator (HO) basis in Jacobi coordinates.
The special properties of HO states allow an
exact antisymmetrization of the basis~\cite{sus}. Our calculations are based on realistic
chiral nucleon-nucleon (NN) interactions
up to fifth order~\cite{N5LO} and YN ones up to second order~\cite{Heiden},~\cite{Heiden2} in the chiral
expansion. The similarity Renormalization Group (SRG) is applied
to NN  and YN interactions to speed up the convergence. We showed our first results
for the $\Lambda$ separation energies $E_b$  of  $^{4}_{\Lambda}{\mbox{He}}$, $^{5}_{\Lambda}{\mbox{He}}$ and
 $^{7}_{\Lambda}{\mbox{Li}}$. We found that in both orders, zeroth (LO) and second (NLO), YN interactions with
with SRG strongly overbind the light hypernuclei. NLO order in general leads to smaller separation energies than
LO. Furthermore, the SRG for both YN interactions induces sizable higher-body (YNN, or even YNNN)
forces. Similar results for LO order are also obtained in~\cite{Rob}.

 The impacts of NN interactions  on the separation energy $E_b$
are also discussed in  detail. The effect of NN forces and NN-SRG on $E_b$ is
negligible for  $^{4}_{\Lambda}{\mbox{He}}$ and $^{5}_{\Lambda}{\mbox{He}}$, however it can be noticeable for
 $^{7}_{\Lambda}{\mbox{Li}}$. We also found that $E_b$ is not strongly dependent on details
of the core structure in  $^{4}_{\Lambda}{\mbox{He}}$ and $^{5}_{\Lambda}{\mbox{He}}$  but somewhat more
 in $^{7}_{\Lambda}{\mbox{Li}}$.

Our first results show the feasibility of Jacobi-NCSM calculations for hypernuclei.  It will be very
interesting to extend the J-NCSM to more complex single-strangeness systems as well as double-strangeness
 hypernuclei. \\[0.3cm]

{\noindent}
{\bf{Acknowledgments}}  This work is supported by DFG and NSFC through funds provided to the
Sino-German CRC 110  ``Symmetries and the
Emergence of Structure in QCD" (grant No. DBC01241). The numerical calculations have been performed
on JUQUEEN and JURECA of the JSC, J\"ulich, Germany.

\setlength{\bibsep}{0.0em}
\begin{thebibliographynotitle}{99}
\bibitem{sus} S.~Liebig, U.-G.~Mei{\ss}ner, A.~Nogga,   Eur. Phys. J. A 52  (2016) 103.
\bibitem{N5LO} E.~Epelbaum, H.~Krebs and U.-G.~Mei{\ss}ner, Eur. Phys. J. A 51  (2015) no.5 53.
\bibitem{Heiden}  H.~Polinder, J.~Heidenbauer., U.-G.~Mei{\ss}ner, Nucl. Phys. A 779 (2006) 244.
\bibitem{Heiden2}  J.~Heidenbauer, S.~Petschauer., N.~Kaiser, U.-G.~Mei{\ss}ner, A.~Nogga, W.~Weise,
 Nucl. Phys. A  915 (2013) 24.
\bibitem{Rob}  R.~Wirth, R.~Roth. {\it et al.}  Phys. Rev. Lett. 117 (2016) 182501.
\end{thebibliographynotitle}

\addcontentsline{toc}{section}{K.~Topolnicki et al.: ${\rm Operator~form~of~nucleon{-}deuteron~scattering}$}

\newabstract 
\begin{center}
{\large\bf Operator form of nucleon-deuteron scattering}\\[0.5cm]
        {\bf Kacper Topolnicki}$^1$,  Jacek Golak$^1$,  Roman Skibi{\'n}ski$^1$,
        Yuriy Volkotrub$^1$, and~Henryk Wita{\l}a$^1$\\[0.3cm]
$^1$M. Smoluchowski Institute of Physics, Jagiellonian University,\\
PL-30348, Krak{\'o}w, Poland\\[0.3cm]
\end{center}

The so-called ``three-dimensional" approach is an
alternative to partial wave calculations with applications
in few-nucleon physics.
It was successfully applied to various problems e.g. two-nucleon scattering
\cite{3df11},~\cite{3df1} or deuteron \cite{3df11} and triton \cite{3df2} bound state calculations.
It was also used to calculate observables in processes that involve
electro-weak probes like
electron \cite{3df22} and muon \cite{muon} induced deuteron disintegrations.
Recently we
performed ``three-dimensional" calculations of neutron-deuteron scattering observables, in both the
elastic and breakup channels, using
an approximate solution of the Faddeev equation \cite{3df3}.
We compared
partial wave and ``three-dimensional" results and concluded that the partial
wave observables converge slowly for certain kinematical configurations of the
deuteron breakup at higher
energies. This motivates us to construct complete ``three-dimensional", three-nucleon
scattering calculations for the full solution of the Faddeev equation.

In order to perform practical calculations related to nucleon-deuteron
scattering without resorting to partial wave
decomposition we developed a general operator form of the scattering amplitude
$\BA{\VC{p} \VC{q}} \OP{T} \KT{\phi}$ where $\OP{T}$ is the three-nucleon
transition operator, $\KT{\phi}$ is an initial product state composed from a
deuteron and a free nucleon, and $\VC{p}, \VC{q}$ are Jacobi momenta
in the final state.
This amplitude is an element of the Faddeev equation:
\[
        \OP{T} \KT{\phi} = \OP{t} \OP{P} \KT{\phi} + \OP{t} \OP{G}_{0} \OP{P} \OP{T}
        \KT{\phi},
\]
where $\OP{P}$ is a permutation operator and $\OP{t}$ is the two-nucleon
transition operator satisfying the Lippmann-Schwinger equation. It has the following general operator form \cite{3d3n}:
\begin{equation} 
        \BA{\VC{p} \VC{q}} \OP{T} \KT{\phi} = \sum_{\gamma^{3N}} \sum_{r = 1}^{64} \tau_{r}^{\gamma^{3N}}(\VC{p} , \VC{q} , \VC{q}_{0})\\
        \KT{\gamma^{3N}} \otimes
        \left(\OP{O}_{r}(\VC{p} , \VC{q} , \VC{q}_{0}) \KT{\alpha} \right)
        \label{general}
\end{equation}
with $\tau_{r}^{\gamma^{3N}}(\VC{p} , \VC{q} , \VC{q}_{0})$ being scalar
functions of the two Jacobi momenta in the final state and the free nucleon
momentum $\VC{q}_{0}$, $\OP{O}_{r}(\VC{p} , \VC{q} , \VC{q}_{0})$ being one of
$64$
spin operators \cite{3d3n} and $\KT{\gamma^{3N}}$ being one of the possible three-nucleon isospin
states.

Using the general form (\ref{general}), the scattering amplitude is defined by
the set of scalar functions $\tau_{r}^{\gamma^{3N}}(\VC{p} , \VC{q} ,
\VC{q}_{0})$. Rewriting the Faddeev equation using (\ref{general}) leads to a significant reduction of numerical complexity
and we hope that it
will allow us to construct full ``three-dimensional", three-nucleon scattering calculations.
This will extend the applicability of the ``three-dimensional" formalism to test three- and
many-body nuclear forces derived from chiral
effective field theory \cite{CHIRAL1},~\cite{CHIRAL2},~\cite{CHIRAL3},~\cite{CHIRAL4}.

{\it Acknowledgements:} This work was supported by the Polish National Science Center under grants No. DEC-2016/21/D/ST2/01120 and DEC-2013/10/M/ST2/ 00420.
Some numerical calculations were performed on the supercomputing clusters of the JSC, J{\"u}lich, Germany.

\setlength{\bibsep}{0.0em}
\begin{thebibliographynotitle}{99}
        \bibitem{3df11} \bibentry{{J. Golak}, {W. Gl{\"o}ckle}, {R. Skibi{\'n}ski},
                {H. Wita{\l}a}, {D. Rozp{\c{e}}dzik}, {K. Topolnicki},{I. Fachruddin}, {Ch.
                Elster}, {A. Nogga}} {Phys. Rev. C} {81}{034006} {2010}
        \bibitem{3df1} \bibentry{{J. Golak}, {R. Skibi{\'n}ski}, {H. Wita{\l}a},
                {K. Topolnicki}, {W. Gl{\"o}ckle}, {A. Nogga}, {H. Kamada}}
                {Few-Body Syst.} {53} {237} {2012}
        \bibitem{3df2} \bibentry{{J. Golak}, {K. Topolnicki},
                {R. Skibi{\'n}ski}, {W. Gl{\"o}ckle}, {H. Kamada},
                {A. Nogga}}{Few-Body Syst.} {54} {2427} {2013}
        \bibitem{3df22} \bibentry{{K. Topolnicki}, {J. Golak},
                {R. Skibi{\'n}ski}, {A.E. Elmeshneb} , {W. Gl{\"o}ckle}, {A. Nogga} , {H. Kamada}
                }{Few-Body Syst.} {54} {2233} {2013}
        \bibitem{muon} \bibentry{{J. Golak}, {R. Skibi{\'n}ski}, {H. Wita{\l}a}, {K.
                Topolnicki}, {A. E. Elmeshneb}, {H. Kamada}, {A. Nogga}, {L. E.
                Marcucci}}{Phys. Rev. C}{90}{024001}{2014}
        \bibitem{3df3} \bibentry{{K. Topolnicki}, {J. Golak}, {R. Skibi{\'n}ski},
                {H. Wita{\l}a},  {C.A. Bertulani}} {Eur. Phys J. A} {51} {132} {2015}
        \bibitem{3d3n} K. Topolnicki et. al., \emph{in preparation.}
        \bibitem{CHIRAL1} \bibentry{{E. Epelbaum}, {H. Krebs}, {Ulf-G.
                Mei{\ss}ner}} {Eur. Phys. J. A} {51} {53} {2015}
        \bibitem{CHIRAL2} \bibentry{{E. Epelbaum}, {H. Krebs}, {Ulf-G.
                Mei{\ss}ner}} {Phys. Rev. Lett.} {115} {122301} {2015}
        \bibitem{CHIRAL3} \bibentry{{D. R. Entem} , {N. Kaiser} , {R. Machleidt} , {Y.
                Nosyk}} {Phys. Rev. C}
                {92} {064001} {2015}
        \bibitem{CHIRAL4} \bibentry{{H. Krebs}, {A. Gasparyan} , {E. Epelbaum}} {Phys. Rev. C}
                {85} {054006} {2012}
\end{thebibliographynotitle}

\addcontentsline{toc}{section}{K.~Vobig: ${\rm Advances~in~the~In{-}Medium~Similarity~Renormalization~Group}$}
\newabstract 
\begin{center}
{\large\bf Advances in the In-Medium Similarity Renormalization Group}\\[0.5cm]
{\bf Klaus Vobig}$^1$ \\[0.3cm]
$^1$Institut f\"ur Kernphysik, Technische Universit\"at Darmstadt,\\
Schlossgartenstr. 2, 64289 Darmstadt, Germany\\[0.3cm]
\end{center}

With the advent of nuclear interactions derived from chiral EFT and their early successes in the description of p-shell nuclei in the No-Core Shell Model (NCSM) \cite{RoLa11},
the \textit{ab initio} description of medium-mass nuclei with chiral NN+3N interactions moved into the focus of several research groups worldwide.

A very successful approach is the In-Medium Similarity Renormalization Group (IM-SRG)
which aims at decoupling an $A$-body reference state $| \Psi_{\text{ref}} \rangle$ from all particle-hole excitations.
This can be achieved via a continuous unitary transformation of the Hamiltonian.
An important aspect of the IM-SRG is that the Hamiltonian is normal ordered with respect to the reference state and typically truncated at the normal-ordered two-body level.
In this way, 3N interactions can be naturally included in a normal-ordered two-body approximation.

A great asset of the IM-SRG is the flexibility and simplicity of its basic concept.
Through different choices of generators and decoupling patterns, the numerical characteristics and efficiency of the methods can be controlled and tailored for specific applications.
The IM-SRG evolved Hamiltonian is easily accessible, hermitian, in contrast to e.g.\ coupled-cluster approaches, and can be used for subsequent calculations.

\begin{figure}
        \begin{center}
        \includegraphics[width=.9\textwidth,clip=true]{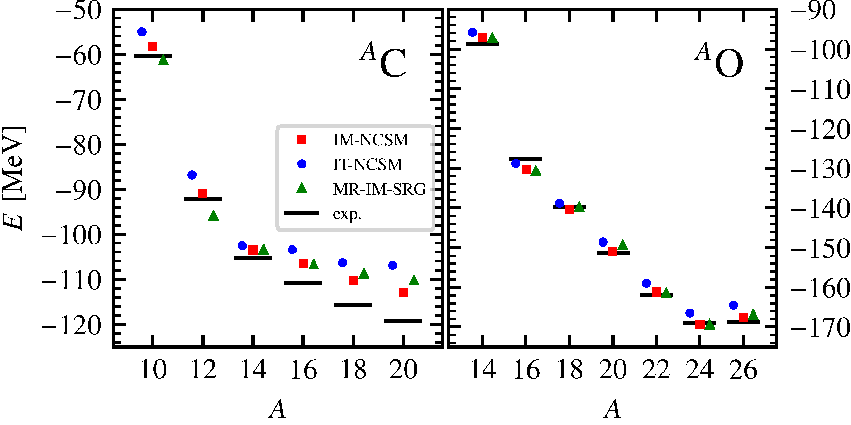}
        \caption{Ground-state energies for the carbon and oxygen chain. Figure taken from \cite{gebrerufael}.}
        \label{fig:o_c_chain}
        \end{center}
\end{figure}

We have first applied this \textit{ab initio} method for the description of ground states of closed-shell nuclei with chiral NN \cite{TsBo11} and NN+3N interactions \cite{HeBo13}.
Going beyond the closed-shell or single-reference version we have generalized the IM-SRG to an open-shell or multi-reference formulation \cite{hergert2014}.
This multi-reference IM-SRG is based on a multi-reference version of normal ordering and Wick's theorem proposed by Kutzelnigg and Mukherjee in quantum chemistry \cite{Kutzelnigg}.
Furthermore, we have merged the IM-SRG with the NCSM leading to the IM-NCSM.
This novel \textit{ab initio} many-body approach uses multi-reference IM-SRG evolved operators as input for a subsequent NCSM calculation and, as we have shown in \cite{gebrerufael}
comes along with a great improvement of the model-space convergence of the subsequent NCSM calculation.
In Fig.\ref{fig:o_c_chain} we have used the IM-NCSM for the calculation of ground state energies of the even isotopes of the carbon and oxygen chain.
Thus, we have advanced the IM-SRG to a versatile tool for the calculation of spectra and spectroscopy of medium-mass nuclei, both closed- and open-shell.
We are investigating extensions of this IM-NCSM approach that will allow us to overcome some of its limitations which are related to the choice of the reference state.
Overcoming these limitations would enable us to explore the whole medium-mass regime of the nuclear chart.

In the context of the consistent NN+3N interactions with semi-local regulators up to N$^3$LO developed within the LENPIC collaboration \cite{lenpic2016},
we will apply the IM-SRG for exploring the order-by-order convergence of the chiral EFT expansion in the medium-mass regime for the first time.
We will extract systematic EFT uncertainties for the ground-state and excitation energies as well as radii of medium-mass systems.
Thus, we will be able to give a complete and systematic quantification of theory uncertainties from the chiral interaction up to the many-body method.

\setlength{\bibsep}{0.0em}
\begin{thebibliographynotitle}{99}
\bibitem{RoLa11}{  R.~ Roth, J.~ Langhammer, et al., Phys. Rev. Lett. 107, 072501 (2011).}
\bibitem{Kutzelnigg}{W. Kutzelnigg and D. Mukherjee, J. Chem. Phys. 107, 432 (1997).}
\bibitem{gebrerufael}{E.~ Gebrerufael, K.~ Vobig, et al., Phys. Rev. Lett. 118, 152503 (2017)}
\bibitem{TsBo11}{ K. Tsukiyama, S. K. Bogner, et al., Phys. Rev. Lett. 106, 222502 (2011).}
\bibitem{HeBo13}{  H. Hergert, S. K. Bogner, et al., Phys. Rev. C 87, 034307 (2013).}

\bibitem{lenpic2016}{ S. Binder, A. Calci, et al., Phys. Rev. C 93, 44002 (2016).}
\bibitem{Entem2017}{ D. R. Entem, R. Machleidt, et al., arXiv:1703.05454 (2017).}
\bibitem{ekstroem2015}{ A. Ekstr\"om, G. R. Jansen, et al., Phys. Rev. C 91, 051301 (2015).}

\bibitem{hergert2014}{ H. Hergert, S. K. Bogner, et al., Phys. Rev. C 90, 41302 (2014).}

\end{thebibliographynotitle}

\addcontentsline{toc}{section}{Yu.~Volkotrub et al.: ${\rm The~OPE{-}Gaussian~force~in~elastic~Nd~scattering}$}
\newabstract 

\begin{center}
{\large\bf The OPE-Gaussian force in elastic Nd scattering}\\[0.5cm]

{\bf Yuriy Volkotrub}$^1$,\let\thefootnote\relax\footnote{This work is a part of the LENPIC project.
It was supported by the Polish National Science Center under Grants No. DEC-2013/10/M/ST2/00420.
The numerical calculations have been performed on the supercomputer
cluster of the JSC, J\"ulich, Germany.}
 Roman Skibi\'nski$^1$, Jacek Golak$^1$, Kacper Topolnicki$^1$, and~Henryk Wita{\l}a$^1$  \\[0.3cm]

$^1$M. Smoluchowski Institute of Physics, Jagiellonian University, 30-384 Krak\'ow, Poland

\end{center}

A reliable estimation of theoretical uncertainties becomes an increasingly important
issue in nuclear physics. 
The most relevant   
sources of such uncertainties are the inaccuracies of the
potential parameters extracted from experimental data,
the uncertainties related to theoretical approaches
and finally the imprecision inherent to numerical methods.
Truncation errors in studies performed with nuclear forces derived within
the chiral effective field theory are an example of the errors arising from the applied theoretical framework.
A prescription to estimate such uncertainties has been
proposed recently in~\cite{imp1} for the two-nucleon (NN) system and in~\cite{Binder} for three-nucleon (3N)
observables.
An estimation of errors stemming from a given numerical scheme is also possible and many
computer science methods can be applied here~\cite{comp_science}. However, it is expected
that numerical computations are performed with a high precision and these types of uncertainties
can be relatively easily minimized.

The propagation of theoretical uncertainties from the NN force to many-body observables is an open question. The recently developed the One-Pion-Exchange-Gaussian (OPE-Gaussian)
potential~\cite{Navarro} delivers a unique opportunity to study this issue. This is because of extensive
attention paid by the authors of Ref.~\cite{Navarro} to determine statistically well defined uncertainties of the
potential parameters.

In this contribution we announce our project for using the OPE-Gaussian force to study
nucleon-deuteron (Nd) scattering and for investigating the uncertainties of 3N observables
stemming from uncertainties of the potential parameters. At the current stage  
we have already obtained predictions for the Nd scattering observables using the central values of the
interaction parameters.

The OPE-Gaussian force is a phenomenological potential, which employs the set of operators
used in the AV18 potential~\cite{AV18} with few additional extensions. It can be decomposed as
\begin{equation}
\label{label1}
V(\vec{r}) = V_{short}(r)\theta (r_{c}-r)+V_{long}(r)\theta (r-r_{c}),
\end{equation}
where $r_{c}$=3~fm and the $V_{long}$ part is just the one-pion exchange force supplemented by the electromagnetic corrections
in the case of the proton-proton force.
The short-range force component can be written as
\begin{equation}
\label{label2}
V_{short}(\vec{r})= \sum\limits_{n=1}^{21} \hat{O}_{n}\left[  \sum\limits_{i=1}^{N} V_{i,n} F_{i,n}(r)\right], 
\end{equation}
where $\hat{O}_{n}$ are the operators from the extended AV18 basis~\cite{AV18}.
The radial Gaussian functions $F_{i,n}(r)$ depend on free parameters $a_{i,n}$ which together with
the $V_{i,n}$ strength parameters are fixed from the NN data. It is worth mentioning that
authors of Ref.~\cite{Navarro} carefully analyzed the existing NN data and constructed
the "3$\sigma$ self-consistent" database which was necessary to obtain
statistical uncertainties of the free potential parameters.

\begin{figure}[htb]  
    \begin{center}
       \includegraphics[scale=0.5,clip=true]{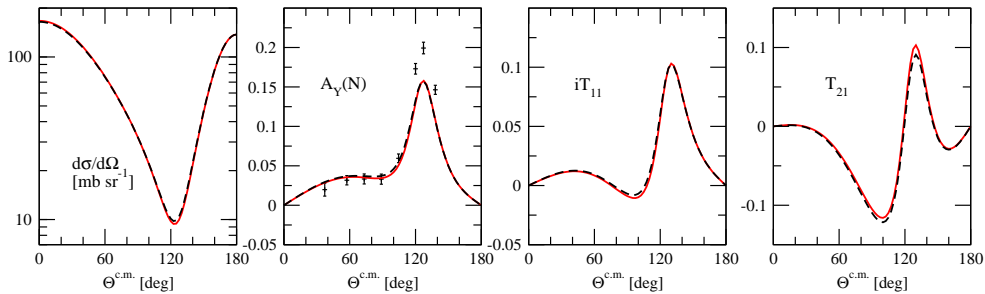}

  \end{center}
   \caption{The differential cross section $\frac{d\sigma}{d\Omega}$, the nucleon analyzing power A$_{\rm Y}$(N) and
the deuteron analyzing powers iT$_{11}$ and T$_{21}$ for the
Nd scattering at E=13 MeV. The black dashed (the red solid) curve represents the AV18 (the OPE-Gaussian) predictions.
Data for the A$_{\rm Y}$(N) are from Ref.~\cite{Cub_ayn13}.
}
\label{fig11}
\end{figure}

Working within the formalism of the Faddeev equation~\cite{G1} we studied
Nd elastic scattering up to incoming nucleon laboratory energy E=200~MeV.
We confirm good behaviour of the new potential and exemplify this in Fig.~\ref{fig11}
for E=13 MeV. We compare there predictions for various observables, obtained with the OPE-Gaussian force
or with the AV18 interaction, and observe very good agreement between predictions based on both models.
This is a starting point for our plans to estimate theoretical uncertainties for these observables.

\vspace{-1.2cm}
\setlength{\bibsep}{0.0em}
\begin{thebibliographynotitle}{99}
\bibitem{imp1} E.Epelbaum, H.Krebs, and Ulf-G.Mei{\ss}ner, Eur. Phys. J. {\bf A51}, 53 (2015).
\bibitem{Binder} S.Binder, {\it et al}., Phys. Rev. {\bf C93}, 044002 (2016).
\bibitem{comp_science} M.L.Overton, Numerical Computing with IEEE Floating Point Arithmetic, SIAM, Philadelphia, 2001.
\bibitem{Navarro} R.Navarro P\'erez, J.E.Amaro, and E.Ruiz Arriola, Phys. Rev. {\bf C89}, 064006 (2014).
\bibitem{AV18} R.B.Wiringa, V.G.J.Stoks, and R.Schiavilla, Phys. Rev. {\bf C51}, 38 (1995).
\bibitem{G1} W.Gl\"ockle  {\it et al.}, Phys. Rept. {\bf 274}, 107 (1996).
\bibitem{Cub_ayn13} J.Cub, {\it et al.}, Few-Body Syst. {\bf 6}, 151 (1989).

\end{thebibliographynotitle}

\end{document}